\documentclass[aps,preprint,amsmath,amssymb]{revtex4}
\usepackage{graphicx,color}
\usepackage{subfigure}

\def\be{\begin{eqnarray}}
\def\ed{\end{eqnarray}}
\def\non{\nonumber}
\def\bbeta{\boldsymbol \eta}
\def\ga{\gamma}
\def\tY{\tilde Y}
\def\Hp{H^\pm}
\def\la{\langle}
\def\ra{\rangle}

\begin{document}

\title{\Large \bf A light charged Higgs boson in two-Higgs doublet model for CDF $Wjj$ anomaly}

\author{Chuan-Hung Chen$^{1,2}$, Cheng-Wei Chiang$^{3,4,2,5}$, Takaaki Nomura$^{3}$, Fu-Sheng Yu$^6$}
\affiliation{
$^{1}$Department of Physics, National Cheng-Kung University, Tainan 701,
Republic of China \\
$^{2}$Physics Division, National Center for Theoretical Sciences, Hsinchu 300,
Republic of China \\
$^{3}$Department of Physics and Center for Mathematics and Theoretical Physics, \\
National Central University, Chungli, Taiwan 32001, Republic of China \\
$^{4}$Institute of Physics, Academia Sinica, Taipei 11925, Republic of China \\
$^{5}$Department of Physics, University of Wisconsin-Madison, Madison, WI 53706, USA \\
$^{6}$Institute of High Energy Physics and Theoretical Physics
Center for Science  Facilities, Chinese Academy of Sciences, Beijing
100049, PeopleÕs Republic of China }

\date{\today}% It is always \today, today,

\begin{abstract}
Motivated by recent anomalous CDF data on $Wjj$ events, we study a possible explanation within the framework of the two-Higgs doublet model.  We find that a charged Higgs boson of mass $\sim$ 140 GeV with appropriate couplings can account for the observed excess.  In addition, we
consider the flavor-changing neutral current effects induced at loop level by the charged Higgs boson on the $B$ meson system to further constrain the model.  Our study shows that the like-sign charge asymmetry $A_{s\ell}^b$ can be of ${\cal O}(10^{-3})$ in this scenario.
\end{abstract}

\maketitle %

%%%%%%%%%%%%%%%%%%%%%%%%%%%%%%%%%%%%%%%%%%%%%%%

Recently the CDF Collaboration reported data indicating an excess of $Wjj$ events where $W$ decayed leptonically \cite{Aaltonen:2011mk}.  The excess shows up as a broad bump between about 120 and 160 GeV in the distribution of dijet invariant mass $M_{jj}$.  This dijet peak can be attributed to a resonance of mass in that range, and the estimated production cross section times the dijet branching ratio is about 4 pb.  However, no statistically significant deviation from the standard model (SM) background is found for $Zjj$ events.  Events with b-jets in the excess region have been checked to be consistent with background.  Moreover, the distribution of the invariant mass of the $\ell\nu jj$ system in the $M_{jj}$ range of 120 to 160 GeV has been examined and indicates no evidence of a resonance or quasi-resonant behavior.  The D\O~Collaboration also performed a similar analysis, but found no excess $Wjj$ events \cite{Abazov:2011af}.  While waiting for further confirmation from the Large Hadron Collider at CERN for the result of either experiment, it is nevertheless worth pursuing the cause of the anomaly observed by CDF.

Many papers have discussed different possible explanations for the excess \cite{Isidori:2011dp,Buckley:2011vc,Yu:2011cw,Eichten:2011sh, Wang:2011uq, Cheung:2011zt, Kilic:2011sr,
AguilarSaavedra:2011zy,Nelson:2011us,He:2011ss,Sato:2011ui,Wang:2011ta, Anchordoqui:2011ag, Dobrescu:2011px,Jung:2011ua,Buckley:2011vs,Zhu:2011ww,Sullivan:2011hu,Ko:2011ns,Plehn:2011nx,
Jung:2011ue,Chang:2011wj,Nielsen:2011wz,Cao:2011yt,Babu:2011yw,Dutta:2011kh,Huang:2011ph,Kim:2011xv,Carpenter:2011yj,Segre:2011ka,Bhattacherjee:2011yh}.
Most of them try to explain the excess by introducing one or more additional new physics particles.  Some consider contributions from exchanging vector bosons, such as $Z'$ and/or $W'$ bosons \cite{Buckley:2011vc, Yu:2011cw,Wang:2011uq, Cheung:2011zt,AguilarSaavedra:2011zy,Wang:2011ta,Anchordoqui:2011ag, Jung:2011ua,Ko:2011ns,Buckley:2011vs,Chang:2011wj,Kim:2011xv,Huang:2011ph}, and neutral color-singlet vector boson \cite{Jung:2011ue}.
Some others analyze the anomaly considering scalar bosons, such as
technipion \cite{Eichten:2011sh} (including technirho), super-partners of fermions \cite{Isidori:2011dp,Kilic:2011sr,Sato:2011ui} (fermions in SUSY model are also considered in Refs.~\cite{Isidori:2011dp,Sato:2011ui}), color octet scalar \cite{Dobrescu:2011px, Carpenter:2011yj}, scalars with flavor symmetry \cite{Nelson:2011us,Zhu:2011ww,Babu:2011yw}, radion
\cite{Bhattacherjee:2011yh}, scalar doublet with no vacuum expectation value (VEV)
\cite{Segre:2011ka}, and new Higgs bosons \cite{Cao:2011yt,Dutta:2011kh}. In Ref.~\cite{Dutta:2011kh}, the two-Higgs doublet model (THDM) is discussed with flavor-changing neutral current (FCNC) interactions allowed through neutral Higgs boson ($H^0$ and $A^0$) exchanges.  Their result favors a light charged Higgs boson.  However, allowing large Yukawa couplings
to leptons in their work has the problem that lepton pairs will be copiously produced, which is not the case in the CDF data.  There are also attempts to explain this puzzle within SM
\cite{He:2011ss,Sullivan:2011hu,Plehn:2011nx,Nielsen:2011wz}.

\begin{figure}[thpb]
\includegraphics*[width=6 in]{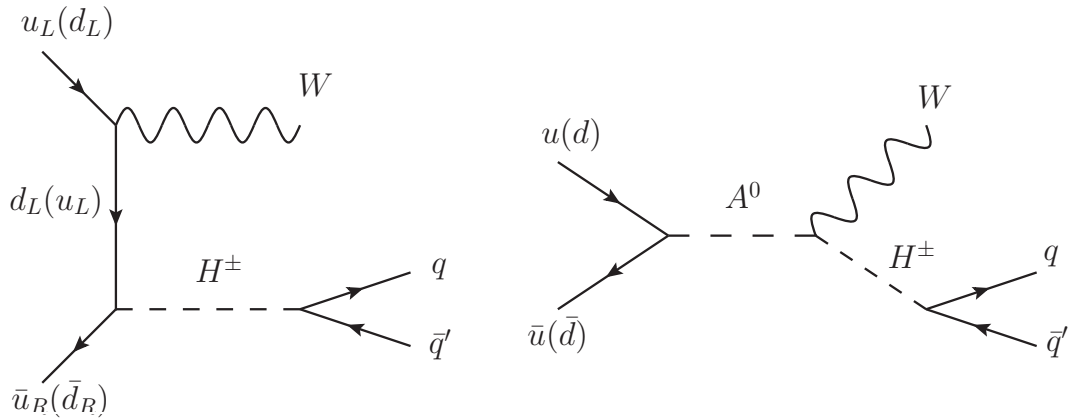}
\caption{Diagrams contributing to the $Wjj$ events in the two-Higgs doublet model.}
\label{fig:HA}
\end{figure}

In this letter, we explore another scenario in the THDM as an explanation.   The fact that the excess dijets are non-b-jets suggests that the new resonance may not couple universally to quarks.  A scalar particle can accommodate this feature more easily than a gauge particle.  We show in Fig.~\ref{fig:HA} two processes in the THDM that can possibly contribute to the excess events.  The dijets come from the decay of the charged Higgs boson $H^\pm$.  Since the CDF Collaboration does not observe any resonance in the invariant mass spectrum of $\ell\nu jj$
for the excess events, we require that the mass of the pseudoscalar Higgs boson $A^0$ to be sufficiently high.  In this case, only Fig.~\ref{fig:HA}(a) is dominant, with the mass of the charged Higgs boson $m_{H^\pm} \sim 140$ GeV, as suggested by data.  Moreover, we assume that the width of $H^\pm$ is sufficiently small in comparison with the jet energy resolution of the experiment.  We note that this is only one possible scenario in the model.  Another scenario is
that $H^\pm$ and $A^0$ are interchanged in Fig.~\ref{fig:HA}, and so are their masses.  We also note in passing that the assumed mass of $\sim 140$ GeV for $H^\pm$ or $A_0$ is consistent with the lower bounds of $76.6$ GeV for $H^\pm$ \cite{Abbiendi:2008be} and 65 GeV for $A^0$ \cite{Abbiendi:2004gn} from LEP experiments.

%%%%%%%%%%%%%%%%%%%%%%%%%%%%%%%%%%%%%%%%%%%%%%%%%%%%%%%%%%%%%%%%%%%%%%%%%
%%%%%%%%%%%%%%%%%%%%%%%%%%%%%%%%%%%%%%%%%%%%%%%%%%%%%%%%%%%%%%%%%%%%%%%%%

%%%%%%%%%%%%%%%%%%%%%%%%%%%%%%%%%%%%%%%%%%%%%%%%%%%%%%%%%%%%%%%%%%%%%%%

The Yukawa sector of the THDM is given by
 \be
-{\cal L}_Y &=& \bar Q_L Y^U_1 U_R \tilde H_1 + \bar Q_L Y^{U}_2 U_R
\tilde H_2  \non
\\
&+& \bar Q_L Y^D_1 D_R H_1 + \bar Q_L Y^{D}_2 D_R H_2+ h.c. ~,
\label{eq:Yu}
 \ed
where $\tilde H_{1,2}$ are two Higgs doublet fields, $\tilde H_k=i\tau_2 H^*_k$, $Q_L$ represents left-handed quark doublets, $U_R$ and $D_R$ are respectively right-handed up-type and down-type quarks, and $Y^{U,D}_{1,2}$ are Yukawa couplings.  Here we have suppressed the generation indices.  The fields $H_1$ and $H_2$ can be rotated so that only one of the two Higgs doublets develops a VEV.  Accordingly, the new doublets are
expressed by
 \be
 h &=&\sin\beta H_1 + \cos\beta H_2 = \left(
            \begin{array}{c}
              G^+ \\
              (v+h^0 +i G^0)/\sqrt{2} \\
            \end{array}
          \right) ~, \non \\
 H &=& \cos\beta H_{1} - \sin\beta H_2=\left(
            \begin{array}{c}
              H^+ \\
              (H^0 +i A^0)/\sqrt{2} \\
            \end{array}
          \right) ~,
 \ed
where $\sin\beta=v_1/v$, $\cos\beta=v_2/v$, $v=\sqrt{v^2_1 + v^2_2}$, $\langle H \rangle = 0$,  $\langle h \rangle = v/\sqrt{2}$.  In our scenario, we assume that $H^\pm$ has mass $\sim 140$ GeV and is responsible for the excess $Wjj$ events observed by CDF.  As a result, Eq.~(\ref{eq:Yu}) can be rewritten as
 \be
 -{\cal L}_{Y} &=& \bar Q_L  Y^U U_R \tilde h + \bar Q_L  Y^D D_R  h \non\\
 &+& \bar Q_L \tY^U U_R \tilde H + \bar Q_L \tY^D_2 D_R H ~
 \label{eq:Yu2}
 \ed
with
 \be
  Y^{F} &=& \sin\beta Y^{F}_1 + \cos\beta Y^{F}_2 \,, \non\\
 \tY^{F} &=& \cos\beta Y^{F}_1 - \sin\beta
 Y^{F}_2\,,
 \ed
and $F=U, D$.  Here, $Y^{F}$ is proportional to the quark mass matrix while $\tY^{F}$ gives the couplings between the heavier neutral and charged Higgs bosons and the quarks.  Clearly, if $Y^{F}$ and $\tY^{F}$ cannot be diagonalized simultaneously, flavor-changing neutral currents (FCNC's) will be induced at tree level and associated with the doublet $H$.  If we impose some symmetry to suppress the tree-level FCNC's, as in type-II THDM, the couplings of the new Higgs bosons are always proportional to the quark masses.  In this case, the excess $Wjj$ events should be mostly b-flavored, which is against the observation.  To avoid this problem, instead of imposing symmetry, we find that $Y^F$ and $\tY^F$ can be simultaneously diagonalized if they are related by some transformation.

To illustrate the desired relationship between $Y^F$ and $\tY^F$, we first introduce unitary matrices $V^F_{L,R}$ to diagonalize $Y^F$ in the following bi-unitary way:
\be
 Y^{\rm dia}_{F} &=&  V^F_L Y^F V^{F\dagger}_{R}
 = \mbox{diag}(Y^1_F, Y^2_F, Y^3_F) ~.
 \label{eq:mF}
 \ed
Using
 \be
 I_{F}= \left(
         \begin{array}{ccc}
           0 & 0 & a \\
           0 & b & 0 \\
           c & 0 & 0 \\
         \end{array}
       \right)\,,
 \label{eq:Imatrx}
 \ed
where $a$, $b$ and $c$ are arbitrary complex numbers, one can easily see that
\be
   \tY^{\rm dia}_F=I_F Y^{\rm dia}_F I^T_F= \left(
         \begin{array}{ccc}
           a^2 Y^3_{F} & 0 & 0 \\
           0 & b^2 Y^2_{F} & 0 \\
           0 & 0 & c^2 Y^1_{F} \\
         \end{array}\right) \label{eq:mFij}
 \ed
is still diagonal.  Now if $\tY^{F}$ and $Y^{F}$ are related by
 \be
\tY^{F} = \bar I^{F}_{L} Y^{F}  \tilde{ I}^{F}_{R} ~,
 \ed
where $\bar I^{F}_{L} = V^{F^\dagger}_{L} I_F V^F_L$ and $\tilde I ^F_{R} = V^{F\dagger}_{R} I^T_F  V^{F}_{R}$, then $\tY^{F}$ and $Y^{F}$ can both be diagonalized by $V^F_{L, R}$, as can be explicitly checked using Eq.~(\ref{eq:mF}) and the unitarity of $V^F_{L(R)}$.  We note that the matrix $I_F$ in Eq.~(\ref{eq:Imatrx}) is not unique.  More complicated examples can be found in Ref.~\cite{Ahn:2010zza}.  Now if the quark mass hierarchy is such that $Y^1_F \ll Y^2_F \ll Y^3_F$, we see in Eq.~(\ref{eq:mFij}) that the hierarchy pattern in $\tY^F$ can be inverted with suitable choices of $a, b$ and $c$.  We note that since $a, b$ and $c$ are arbitrary complex numbers, all elements in $\tY^F$ are also complex in general.  As a result, the couplings between the $H$ doublet and light quarks are not suppressed by their masses.  Moreover, the coupling to $b$ quarks can be suppressed.

To proceed the analysis, we write down the relevant interactions in terms of physical eigenstates:
 \be
 -{\cal L}_{H^{\pm}, H^0, A^0} &=& \left( \bar u_R  \bbeta^{U^\dagger} u_L + \bar d_L \bbeta^D d_R\right) \frac{H^0 + i A^0}{\sqrt{2}} \non \\
 &+& \left( - \bar u_R \bbeta^{U^\dagger} {\bf V} d_L + \bar u_L {\bf V} \bbeta^D d_R \right) H^+ + h.c.\,, %\non \\
  \label{eq:int_H}
 \ed
where ${\bf V}$ is the Cabibbo-Kobayashi-Maskawa (CKM) matrix, and $\bbeta^{F}=$ diag$(\eta^F_1, \eta^F_2, \eta^F_3)$ contains three free parameters.  For simplicity and illustration purposes, we will consider two schemes:
\begin{eqnarray}
\mbox{(I): } &&
\eta_i \equiv \eta_i^U = \eta_i^D ~\mbox{ with } i = 1,2,3 ~; \\
\mbox{(II): } &&
\eta^U \equiv \eta_i^U ~\mbox{ and }~
\eta^D \equiv \eta_j^D ~\mbox{ for } i = 1,2,3 \mbox{ and } j=1, 2 ~.
\end{eqnarray}
To suppress the coupling with the $b$ quark, we require that $\eta_3 \ll \eta_{1,2}$ in Scheme (I) or $\eta^D_3 \ll 1$ in Scheme (II).

In either scheme, we search for the parameter space that can explain the excess $Wjj$ events, subject to the constraint $\sigma_{Wjj} \equiv \sigma(p \bar{p} \rightarrow W H^\pm) BR(H^\pm \rightarrow jj) = 4$ pb, as observed by CDF.  Moreover, we consider a $25\%$ uncertainty in the extracted $\sigma_{Wjj}$.  In the scenario of a heavy CP-odd Higgs boson, we ignore the contribution from Fig.~\ref{fig:HA}(b) and consider only the t-channel Feynman diagram.  The contribution of Fig.~\ref{fig:HA}(b) is one order less than that of Fig.~\ref{fig:HA}(a) when $m_{A^0} \agt 650$ GeV.

\begin{figure}[hptb]
\includegraphics*[width=4 in]{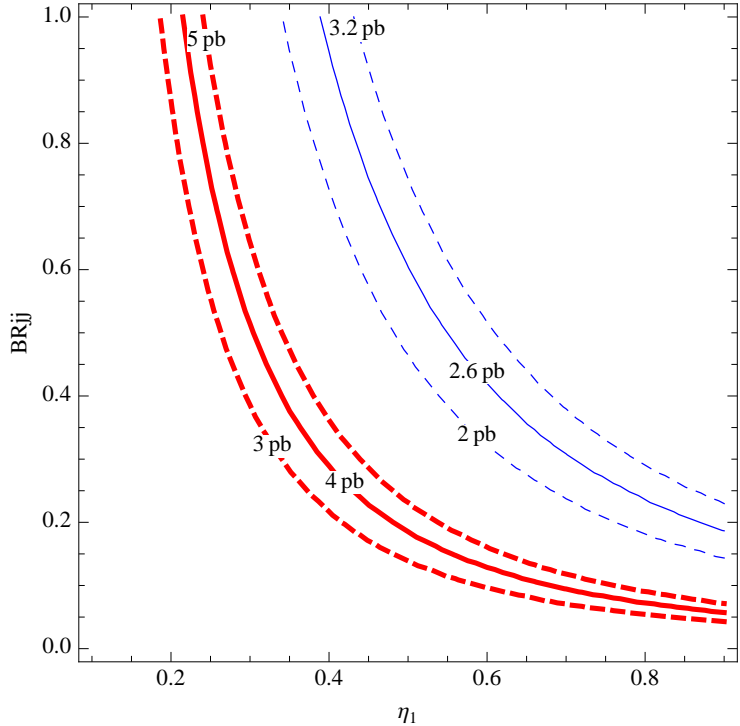}
\caption{Contours of $\sigma_{Wjj} = (4 \pm 1)$ pb (thick red curves) and $\sigma_{Zjj}=(2.6 \pm 0.6 )$ pb (thin blue curves) for Scheme (I).  In this scenario, we take $m_{H^\pm} = 144$ GeV and $A^0$ is sufficiently heavy.  A $K$ factor of 1.3 is used in computing the cross section.}
\label{fig:contour5}
\end{figure}

We first consider Scheme (I).  Due to small parton distribution functions (PDF's) associated with charm and strange quarks in the proton (or anti-proton), we find that $\eta_2$ does not play a significant role in determining $\sigma_{Wjj}$.  Therefore, $\sigma_{Wjj}$ mainly depends on $\eta_1$, the coupling between $H^\pm$ and quarks of the first generation, and the hadronic branching ratio of $H^\pm$, ${\cal B}_{\rm jj} \equiv BR(H^\pm \rightarrow jj)$.  In Fig.~\ref{fig:contour5}, we fix $\eta_2 = 0.1$.  The red curves on the $\eta_1$-${\cal B}_{\rm jj}$ plane are contours corresponding to $\sigma_{Wjj} = (4 \pm 1)$ pb.  In this analysis, we took mass of $m_{H^\pm} = 144$ GeV in accordance with the CDF result \cite{Aaltonen:2011mk}.

In principle, the same t-channel diagram in Fig.~\ref{fig:HA} can contribute to $Zjj$ events.  However, the couplings of the $Z$ boson to charged leptons are more suppressed than $W$.
The blue curves in Fig.~\ref{fig:contour5} are contours of $\sigma_{Zjj} \equiv \sigma(p \bar{p} \rightarrow Z H^\pm) BR(H^\pm \rightarrow jj)$ bring around 2.6 pb, which is the cross section of SM background process $p \bar{p} \rightarrow Z Z+Z W \rightarrow Z jj$.  We see that the preferred parameter region of the red curves have $\sigma_{Zjj}$ well below the SM background.

%%%%%%%%%%%%%%%%%%%%%%%%%%%%%%%%%%%%%%%%%%%%%%%%%%%%%%%%%%%%%%%%%%%%%%%%%
%%%%%%%%%%%%%%%%%%%%%%%%%%%%%%%%%%%%%%%%%%%%%%%%%%%%%%%%%%%%%%%%%%%%%%%%%

Using the extracted parameter space, we then compute the total width of $H^\pm$ using the partial width formula
\begin{equation}
\Gamma_q = \sum_{i,j=1,2}
\frac{3}{16 \pi } m_{H^{\pm}}    |V_{u_i d_j}|^2 [(\eta^U_i)^2+(\eta^D_j)^2]
\end{equation}
and ${\cal B}_{\rm jj}$.  Note that the $b$-quark coupling has been taken to be zero in the above formula.  When ${\cal B}_{\rm jj} \agt 0.8$ for $\eta_1 = \eta_2$ or 0.7 for $\eta_1 \gg \eta_2$, the total width $\Gamma_{H^\pm} \alt 2$ GeV, consistent with our narrow width approximation.  This suggests that the charged Higgs boson couple dominantly to quarks instead of leptons.

\begin{figure}[hptb]
\subfigure{
\includegraphics*[width=3 in]{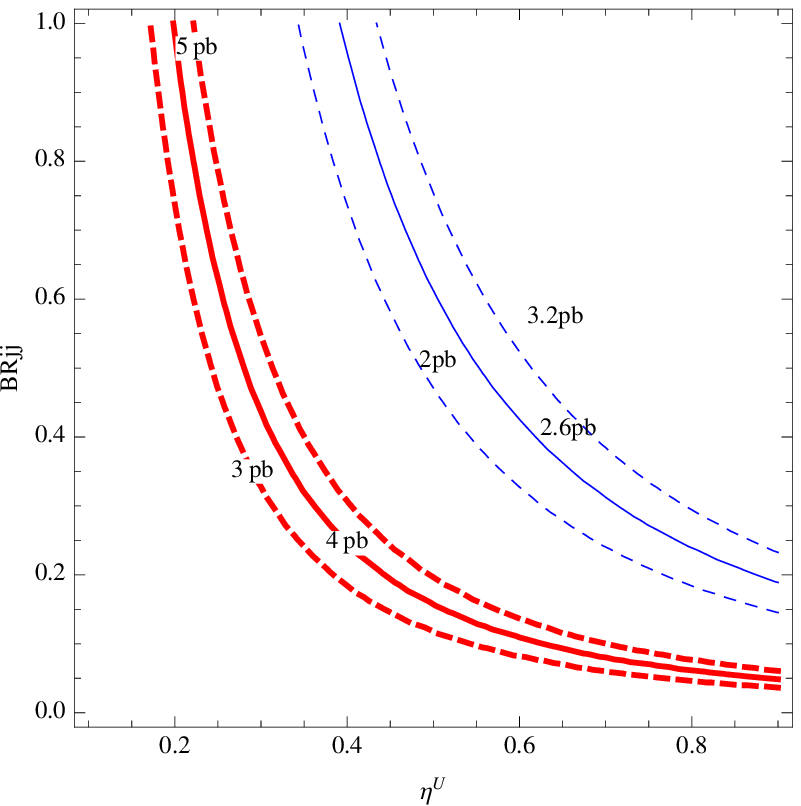}
\label{contour4a}
}
\subfigure{
\includegraphics*[width=3 in]{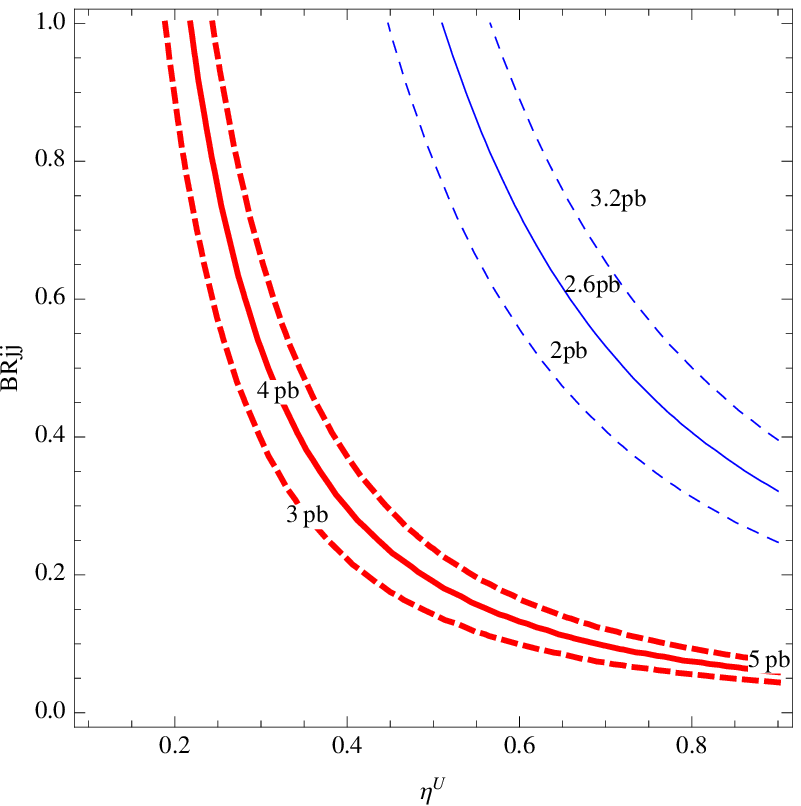}
\label{contour4b}
} \\
(a) \hspace{7cm} (b)
\caption{Same as Fig.~\ref{fig:contour5} but for Scheme (II).  Results for $\eta^D = \eta^U$ are shown in plot (a), and results for $\eta^D=0.1 \eta^U$ are shown in plot (b).  }
\label{fig:contour6}
\end{figure}

We now consider two cases in Scheme (II): (a) $\eta^D = \eta^U$ and (b) $\eta^D=0.1 \eta^U$.  The independent parameters are then $\eta^U$ and ${\cal B}_{\rm jj}$.  Plots in Fig.~\ref{fig:contour6} show that it is preferred to have $\eta^D < \eta^U$ because it helps suppressing $Zjj$ production.  Likewise, when ${\cal B}_{\rm jj} \agt  0.8$, the total width $\Gamma_{H^\pm} \alt 2$ GeV, again consistent with our narrow width approximation.

%%%%%%%%%%%%%%%%%%%%%%%%%%%%%%%%%%%%%%%%%%%%%%%%%%%%%%%%%%%%%%%%%%%%%%%%%
%%%%%%%%%%%%%%%%%%%%%%%%%%%%%%%%%%%%%%%%%%%%%%%%%%%%%%%%%%%%%%%%%%%%%%%%%

\begin{figure}[hptb]
\subfigure{
\includegraphics*[width=3 in]{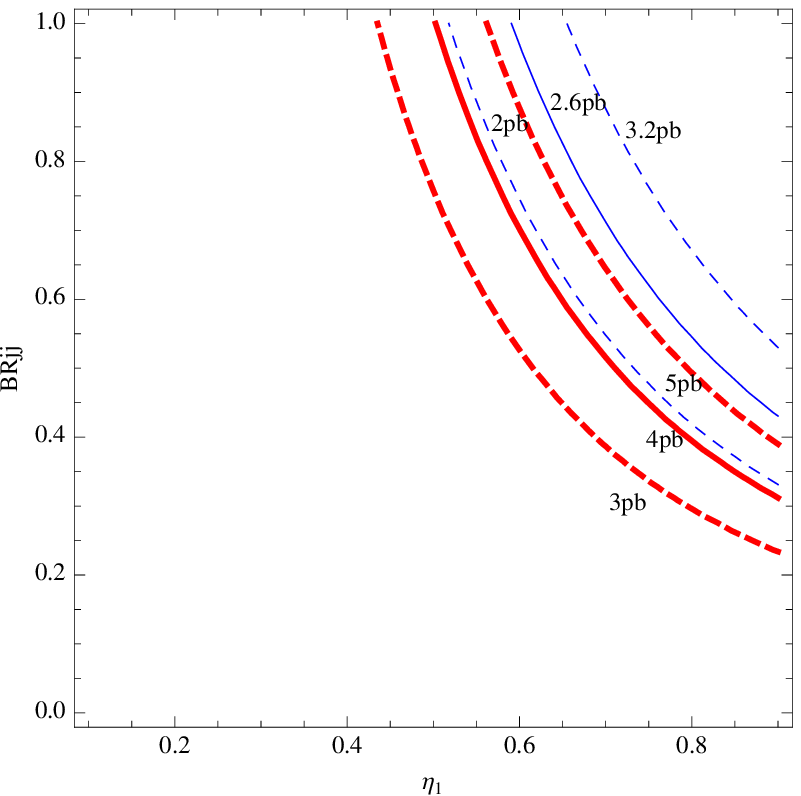}
}
\subfigure{
\includegraphics*[width=3 in]{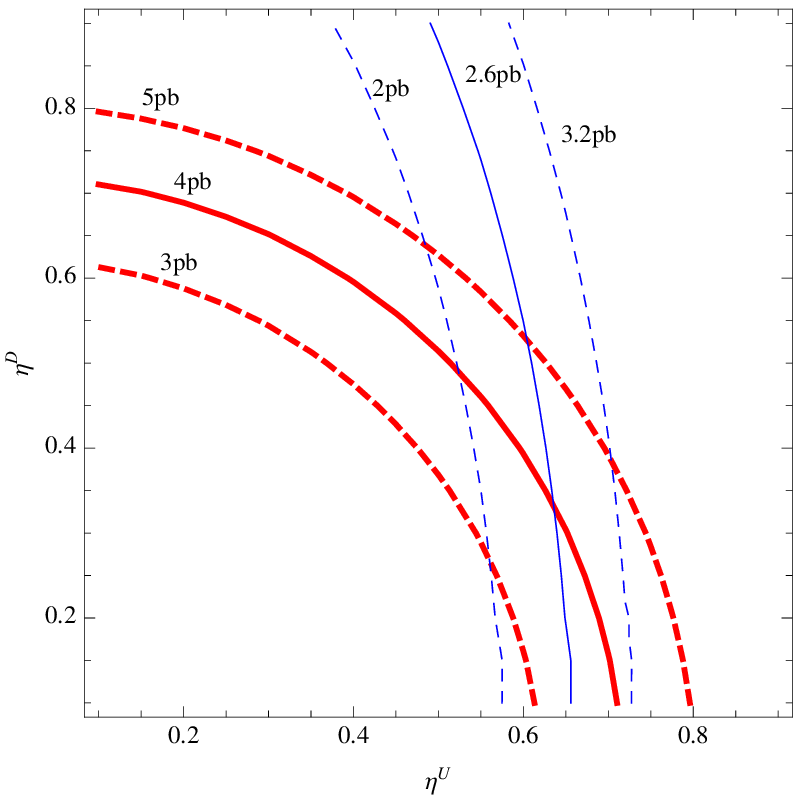}
} \\
(a) \hspace{7cm} (b)
\caption{Contours of $\sigma(p \bar{p} \rightarrow W A^0) Br(A^0 \rightarrow jj) = (4 \pm 1)$ pb (thick red curves) and $\sigma(p \bar{p} \rightarrow Z A^0) Br(A^0 \rightarrow jj) = (2.6 \pm 0.6)$ pb (thin blue curves) for Scheme (I) in plot (a) and Scheme (II) in plot (b).  In this scenario, we take $m_{A^0} = 144$ GeV and $H^\pm$ is sufficiently heavy.  A $K$ factor of 1.3 is used in computing the cross section.}
\label{fig:contour2}
\end{figure}

We note in passing that one can also consider the scenario where the roles of $H^\pm$ and $A^0$ are interchanged, with the former being heavy and the latter having a mass of $144$ GeV.  However, the parameter region for explaining the $Wjj$ events predicts a $Zjj$ rate very close to the SM background in Scheme (I), as shown in Fig.~\ref{fig:contour2}(a).  In Scheme (II), null deviations of $Zjj$ and b-jets disfavor the small and large $\eta^D$ regions, respectively, as shown in Fig.~\ref{fig:contour2}(b).  Therefore, in comparison the previous scenario with light charged Higgs boson and heavy CP-odd Higgs boson is favored.  We will thus exclusively consider such a scenario in the following analysis of low-energy constraints.

%%%%%%%%%%%%%%%%%%%%%%%%%%%%%%%%%%%%%%%%%%%%%%%%%%%%%%%%%%%%%%%%%%%%%%%%%
%%%%%%%%%%%%%%%%%%%%%%%%%%%%%%%%%%%%%%%%%%%%%%%%%%%%%%%%%%%%%%%%%%%%%%%%%

\begin{figure}[thpb]
\includegraphics*[width=4 in]{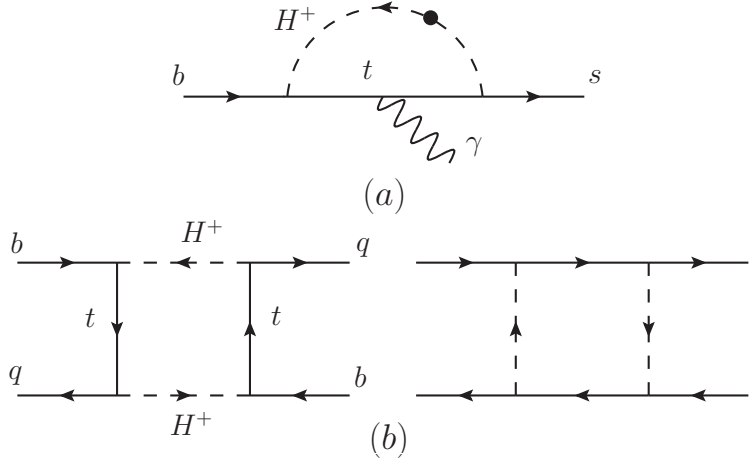}
\caption{ (a) $b\to s \gamma$ transition, where the dot indicates another possible place to attach the photon and (b) $M^q_{12}$ with $q=d, s$ induced by $H^{\pm}$.}
\label{fig:pheno}
\end{figure}

If the charged Higgs boson is a candidate for the new resonance, it will also induce interesting phenomena in low-energy systems, where the same parameters are involved.
We find that the most interesting processes are the $B\to X_s \ga$ decay and the like-sign charged asymmetry (CA) in semileptonic $B_q$ ($q=d,s$) decays.  To simplify our presentation, we leave detailed formulas in Appendix \ref{sec:formulas}.
Using the interactions in Eq.~(\ref{eq:int_H}), the effective Hamiltonians for the $b\to s \ga$ and $\Delta B=2$ processes induced by $H^\pm$, as shown in Fig.~\ref{fig:pheno}, are respectively given by
\be
{\cal H}_{b\to s \ga} &=& \frac{V^*_{ts} V_{tb} |\eta^U_3|^2 }{16m^2_{\Hp}}\left( Q_t I_1(y_t) + I_2(y_t) \right) {\cal O}_{7\ga} \non \\
&+& \frac{V^*_{ts} V_{tb} \eta^{D^*}_2 \eta^{U^*}_{3} }{8m^2_{\Hp}} \frac{m_t}{m_b} \left( Q_t J_1(y_t) + J_2(y_t) \right) {\cal O}'_{7\ga}\,, \non \\
 {\cal H}(\Delta B=2)&=& \frac{\left( V^*_{tq} V_{tb}  |\eta^U_3|^2 \right)^2}{4(4\pi)^2 m^2_{\Hp}} I_3 ( y_t ) \bar q \ga_\mu P_L b \bar q \ga^\mu P_L b\non \\
 &-&  \frac{\left( V^*_{tq} V_{tb} \eta^{D^*}_2 \eta^{U^*}_{3} \right)^2}{2(4\pi)^2 m^2_{\Hp}} y_t J_3 ( y_t ) \left(\bar q P_L b \right)^2 ~,
 \label{eq:EH}
 \ed
where $y_t \equiv m^2_t/m^2_{\Hp}$, ${\cal O}_{7\ga}$ and ${\cal O}'_{7\ga}$ are defined in the Appendix, $Q_t=2/3$ is the top-quark electric charge, and
 \be
 I_1(a) &=& \frac{2+5a-a^2}{6(1-a)^3} + \frac{a\ln a}{(1-a)^4}\,,\non\\
J_1(a)&=&\frac{3-a}{(1-a)^2}+\frac{\ln a}{(1-a)^3}\,, \non \\
I_2(a) &=& \frac{1-5a-2a^2}{6(1-a)^3} - \frac{a^2\ln a}{(1-a)^4}\,,\non\\
J_2(a)&=& \frac{1+a}{2(1-a)^2} +\frac{a\ln a}{(1-a)^3}\,, \non \\
 I_3(a)&=&\frac{1+a}{2(1-a)^2} + \frac{a \ln a}{(1-a)^3} \,, \non \\
 J_3(a)&=& -\frac{2}{(1-a)^2} - \frac{1+a}{(1-a)^3}\ln a\,.
 \ed

%%%%%%%%%%%%%%%%%%%%%%%%%%%%%%%%%%%%%%%%%%%%%%%%%%%%%%%%%%%%%%%%%%%%%%%%%
%%%%%%%%%%%%%%%%%%%%%%%%%%%%%%%%%%%%%%%%%%%%%%%%%%%%%%%%%%%%%%%%%%%%%%%%%

 Using the hadronic matrix elements defined by
 \be
 \la B_q | \bar q \ga_\mu P_L b \bar q\ga^\mu P_L b| \bar B_q \ra &=& \frac{1}{3} m_{B_q} f^2_{B_q} \hat B_q\,, \non \\
  \la B_q | \bar q  P_L b \bar q P_L b| \bar B_q \ra &\approx& - \frac{5}{24}\left( \frac{m^2_{B_q}}{m_b+m_q}\right)^2 m_{B_q} f^2_{B_q} \hat B_q ~,
 \ed
and the formulas given in the Appendix, the dispersive part of $B_q$-$\overline B_q$ mixing is found to be
  \be
  M^{q}_{12} &=& M^{q, SM}_{12} + M^{q, H}_{12} = M^{q, SM}_{12} \Delta_q e^{i\phi^\Delta_q}\,,
  \ed
where
  \be
 \Delta_q &=& \left( 1+ R^{q^2}_H +2 R^q_H \cos2\theta^q_H \right)^{1/2}\,, \non \\
 \theta^q_H &= & arg\left(\frac{M^{q, H}_{12}}{M^{q, SM}_{12}}\right)\,, \ \ \
 R^q_H=\left|\frac{M^{q, H}_{12}}{M^{q, SM}_{12}}\right|^2\, , \non \\
 \tan\phi^\Delta_q &=& \frac{R^q_H \sin2\theta^q_H}{1+R^q_H \cos2\theta^q_H }\,.
 \ed
Since the charged Higgs boson is heavier than the $W$ boson, its influence on $\Gamma^s_{12}$ is expected to be insignificant.  Therefore, we set $\Gamma^q_{12} \approx \Gamma^{q, SM}_{12}$ in our analysis.  Using $\phi_q = \mbox{arg}(-M^q_{12}/\Gamma^q_{12})$, the $H^\pm $-mediated wrong-sign CA defined in Eq.~(\ref{eq:aqsl}) is given by
 \be
 a^q_{s\ell} (H^\pm)&=&
 \frac{1}{\Delta_q} \frac{\sin\phi_q}{\sin\phi^{\rm SM}_{q}}a^{q}_{s\ell}({\rm SM}) ~,
 \label{eq:aqsl_new}
 \ed
with $\phi^{\rm SM}_q=-2\beta_q -\ga^{\rm SM}_{q}$ \cite{Lenz:2011ti} and $\phi_q=\phi^{\rm SM}_{q} + \phi^{\Delta}_{q}$.  Consequently, the like-sign CA in Eq.~(\ref{eq:Absl}) is read as $A^{b}_{s\ell}\approx 0.506\, a^d_{s \ell} (\Hp ) + 0.494\, a^{s}_{s \ell}(\Hp)$, where the SM contributions are  $a^{s}_{s \ell}(\rm SM)\approx 1.9 \times 10^{-5}$ and $a^{d}_{s \ell}(\rm SM) \approx -4.1\times 10^{-4}$ \cite{Lenz:2011ti,Lenz:2006hd}.

\begin{figure}[bthp]
\includegraphics*[width=6 in]{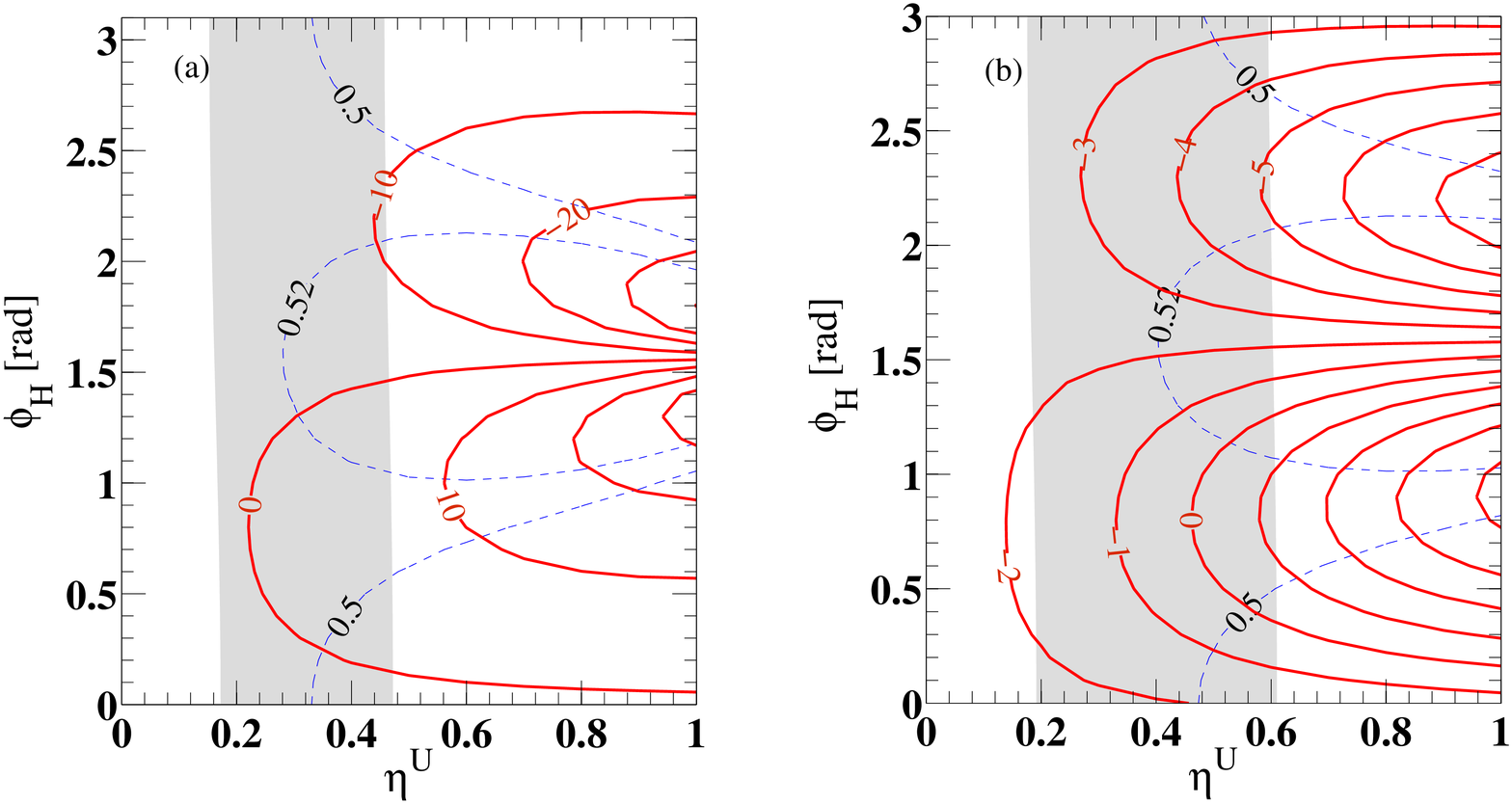} \\
(a) \hspace{7cm} (b)
\caption{Contours of $A^b_{s\ell}$ (in units of $10^{-4}$) on the $\eta^U$-$\phi_H$ plane, where the shaded band and the dashed curves show the constraints from the measured ${\cal B}(B\to X_s \ga)$ and $\Delta m_{B_d}$, respectively, within their $2\sigma$ errors.}
\label{fig:Bprocess}
\end{figure}

In the following, we numerically study the charged Higgs contributions to the $B \to X_s \ga$ decay.  In Scheme (I), as $\eta_3 = \eta^U_3 = \eta^D_3 \ll 1$ is assumed, it is clear that their contributions to the $B$ decay are small. Thus, we concentrate on the analysis of Scheme (II).  Using Eqs.~(\ref{eq:EH}) and (\ref{eq:int_bsga}), the $\Hp$-mediated Wilson coefficients  for $b\to s\ga$ are given by
\be
\delta C_7 &=&-\frac{|\eta^U|^2}{8\sqrt{2} m^2_{\Hp} G_F}
\left( Q_t I_1(y_t) + I_2(y_t) \right)\,, \non \\
\delta C'_7 &=&-\frac{\eta^{D^*} \eta^{U^*}  }{4\sqrt{2} m^2_{\Hp} G_F}  \frac{m_t}{m_b}
\left( Q_t J_1(y_t) + J_2(y_t) \right) \,.
\ed
Here the enhancement mainly comes from the large $\eta^U$ coupling.  Taking Eq.~(\ref{eq:Br_bsga}) and setting $\eta^D=\rho \eta^U$ and $\phi_H=\mbox{arg}(\eta^{D^*} \eta^{U^*})$, one can calculate the branching ratio of $B\to X_s \ga$ as a function of $\eta^U$ and $\phi_H$.  The $2\sigma$ range of experimental measurement ${\cal B}(B\to X_s\ga)=(3.55 \pm  0.26)\times 10^{-4}$ \cite{TheHeavyFlavorAveragingGroup:2010qj} demands the two parameters to be within the shaded bands in Fig.~\ref{fig:Bprocess}, where plot (a) and (b) use $\rho=1$ and $0.5$, respectively.  The results show that ${\cal B}(B\to X_s \ga)$ is insensitive to the new phase $\phi_H$ and that the allowed range of $\eta^U$ is compatible with the above analysis for the $Wjj$ events.  In addition, we also show in Fig.~\ref{fig:Bprocess}  the constraint from measured $\Delta m_{B_d}$ (dashed blue curves).  We only take into account $\Delta m_{B_d}$ here simply because the measurement $\Delta m_{B_d}=0.507\pm 0.005$ ps$^{-1}$ is more precise and thus stringent than $\Delta m_{B_s}=17.78\pm 0.12$ ps$^{-1}$.  It is observed that the measurement of $\Delta m_{B_d}$ further excludes some of the parameter space allowed by the $B\to X_s \ga$ decay.  Finally, we superimpose contours of the like-sign CA (solid red curves) in Fig.~\ref{fig:Bprocess}.  The like-sign CA has a strong dependence on the value of $\rho$.  When $\rho \sim O(1)$, $A^{b}_{s\ell}$ can be of the order of $10^{-3}$.  However, it drops close to the SM prediction when $\rho\sim O(0.1)$.

We now comment on the constraints from $K$-$\overline K$ and $D$-$\overline D$ mixings.  In the usual THDM, contributions from box diagrams involving the charged Higgs bosons to the mass difference are important because the charged Higgs couplings to quarks are proportional to their masses.  Therefore, the Glashow-Iliopoulos-Maiani (GIM) mechanism \cite{Glashow:1970gm} is not effective to suppress such new physics effects \cite{Abbott:1979dt}.  In the scenarios considered in this work, the $H^\pm qq'$ couplings are simply proportional to the CKM matrix elements.  Therefore, the box diagrams involving the charged Higgs boson will have GIM cancellation in the approximation that the masses of quarks in the first two generations are negligible.  Although the third generation fermions do not have GIM cancellation, the associated CKM matrix elements are much suppressed.  In addition, the new effective operators thus induced will be further suppressed by powers of $m_{W}/m_{\Hp}$.  For example, with $m_{\Hp}=140$ GeV, $\rho=1$, $\eta^U=0.4$ and the dispersive part of $K$-$\overline K$ mixing given in Eq.~(\ref{eq:dk2}), we obtain $\Delta m_K \sim 1.58\times 10^{-17}$ GeV, which is two orders of magnitude smaller than the current measurement, $(\Delta m_K)^{\rm exp}=(3.483\pm 0.006)\times 10^{-15}$ GeV \cite{PDG08}.  We note that unlike the conventional THDM, where the diagrams with one $W^\pm$ and one $H^\pm$ in the loop are important \cite{K}, in Scheme (II) of our model the GIM mechanism is very effective in the massless limit of the first two generations of fermions.  This has to do with the fact that the charged Higgs couplings to these quarks are independent of quark masses.  The contributions from diagrams with the top-quark loop are also negligible due to the suppression of the small CKM matrix elements $(V_{ts}V^*_{td})^2$, as in the conventional THDM with small $\tan\beta$.  The relevant formulas for $K$-$\overline K$ mixing from these contributions are given by Eqs.~(\ref{eq:dk2}) and (\ref{eq:dk2WH}).  The constraint from $D$-$\overline D$ mixing is even weaker in view of current measurements \cite{delAmoSanchez:2010xz} and the fact that new physics contributions are both GIM and doubly Cabibbo suppressed.

%%%%%%%%%%%%%%%%%%%%%%%%%%%%%%%%%%%%%%%%%%%%%%%%%%%%%%%%%%%%%%%%%%%%%%%%%
%%%%%%%%%%%%%%%%%%%%%%%%%%%%%%%%%%%%%%%%%%%%%%%%%%%%%%%%%%%%%%%%%%%%%%%%%

In summary, we have studied a scenario of the two-Higgs doublet model as a possible explanation for the excess $Wjj$ events observed by the CDF Collaboration.  In this scenario, the charged Higgs boson has a mass of about $144$ GeV and decays into the dijets.  We find that both Scheme (I) and Scheme (II) considered in this work can explain the $Wjj$ anomaly while not upsetting the constraints of $Zjj$ and b-jets being consistent with standard model expectations.  When applying the scenario to low-energy $B$ meson phenomena, we find that very little constraint can be imposed on Scheme (I) as $\eta_3$ couplings to the third generation quarks are assumed to be negligible.  Scheme (II), on the other hand, has constraints from the $B \to X_s \ga$ decay and $\Delta m_{B_d}$.  In particular, we find that if $\eta^D$ for the first two generations is of the same order of magnitude as $\eta^U$, it is possible to obtain $A_{s\ell}^b \sim {\cal O}(10^{-3})$.  Constraints from $K$-$\overline K$ and $D$-$\overline D$ mixings are found to be loose primarily due to the GIM cancellation.

\section*{\bf Acknowledgments}

F.~S.~Y.~would like to thank Prof.~Cai-Dian Lu and Dr.~Xiang-Dong
Gao for useful discussions. C.~H.~C.~ was supported by NSC Grant No.
NSC-97-2112-M-006-001-MY3. C.~W.~C and T.~N.~were supported in part
by NSC-97-2112-M-001-004-MY3 and 97-2112-M-008-002-MY3.  The
work of F.~S.~Y.~was supported in part by National Natural Science
Foundation of China under the Grant Nos.~10735080 and 11075168 and
National Basic Research Program of China (973)  No.~2010CB833000.

\begin{appendix}
%-------------------------------
\section{  $B\to X_s \ga$  and like-sign CA in semileptonic $B_q$ decays }\label{sec:formulas}
%-------------------------------

For the $B \to X_s \gamma$ decay, the effective Hamiltonian is
 \be
 {\cal H}_{b\to s\ga} &=& - \frac{G_F}{\sqrt{2}} V^*_{ts} V_{tb} \left ( C_{7}(\mu) O_{7\ga} + C'_{7}(\mu) O'_{7\ga} \right)
  \label{eq:int_bsga}
 \ed
with
 \be
 O_{7\ga} &=& \frac{e m_b}{8\pi^2} \bar s \sigma_{\mu \nu} (1+\ga_5) b F^{\mu \nu} \,, \non \\
 O'_{7\ga}&=& \frac{e m_b}{8\pi^2} \bar s \sigma_{\mu \nu} (1-\ga_5)  b F^{\mu \nu}\,.
 \ed
The branching ratio is given by \cite{bsga}
 \be
 {\cal B}(B\to X_s \ga)_{E_\ga > 1.6 GeV} &=&  \left[  a_{00}  +a_{77} \left( |\delta C_7|^2 + |\delta C'_7|^2 \right)   \right. \non \\
 &+& \left.  a_{07}Re(\delta C_7) + a'_{07} Re (\delta C'_7) \right]\times 10^{-4} ~,
  \label{eq:Br_bsga}
 \ed
where $a_{00}=3.15\pm 0.23$, $a_{07}=-14.81$, $a_{77}=16.68$, and $a'_{07}=-0.23$.  The parameters $\delta C_7 = C^{NP}_7 $ and $\delta C'_7\approx C^{NP}_7$ stand for new physics contributions.

To understand the like-sign CA, we start with a discussion of relevant phenomena.  In the strong interaction eigenbasis, the Hamiltonian for unstable $\bar B_q$ and $B_q$ states
is
 \be
 {\bf H }={\bf  M^q }- i \frac{{\bf \Gamma^q }}{2}\,,
 \ed
where $\bf \Gamma^{q}$ ($\bf M^q$) denotes the absorptive (dispersive) part of the $\overline B_q \leftrightarrow B_q $ transition.  Accordingly, the time-dependent wrong-sign CA in semileptonic $B_q$ decays is defined and given \cite{PDG08} by
 \be
a^q_{s\ell}&\equiv& \frac{\Gamma(\bar B_q(t) \to \ell^+ X)- \Gamma( B_q(t) \to \ell^- X)}{\Gamma(\bar B_q(t) \to \ell^+ X)
+\Gamma( B_q(t) \to \ell^- X)}\,,\non \\
&\approx& {\rm Im}\left( \frac{\Gamma^{q}_{12}}{M^{q}_{12}}\right) \label{eq:aqsl}\,.
 \ed
Here, the assumption of $\Gamma^q_{12}\ll M^q_{12}$ in the $B_q$ system has been used. Intriguingly, $a^q_{s\ell}$ is actually not a time-dependent quantity.  The relation
between the wrong and like-sign CAs is defined and expressed by \cite{Abazov:2010hv,Grossman:2006ce}
 \be
 A^b_{s\ell} &=& \frac{\Gamma(b\bar b\to \ell^+ \ell^+ X) - \Gamma(b\bar b\to
 \ell^- \ell^- X)}{\Gamma(b\bar b\to \ell^+ \ell^+ X) + \Gamma(b\bar b\to \ell^- \ell^- X)}\,,\non\\
 &=& 0.506(43) a^d_{s\ell} + 0.494(43) a^s_{s\ell}\,. \label{eq:Absl}
 \ed
Clearly, the like-sign CA is associated with the wrong-sign CA's of the $B_d$ and $B_s$ systems.  Since the direct measurements of $a^d_{s\ell}$ and $a^s_{s\ell}$ are still quite imprecise, either $b\to  d$ or $b\to s$ transition or both can be the source of the
unexpectedly large $A^b_{s\ell}$ observed experimentally.

In order to explore new physics effects, we parameterize the transition matrix elements as
 \be
M^{q}_{12} &=& M^{q,{\rm SM}}_{12} \Delta^M_q e^{i\phi^{\Delta}_q }\,, \non \\
\Gamma^q_{12} &=& \Gamma^{q, {\rm SM}}_{12} \Delta^{\Gamma}_{q} e^{i\ga^{\Delta}_q} ~,
\label{eq:MG}
\ed
for $q = d, s$, where
 \be
 M^{q, {\rm SM}[{\rm NP}]}_{12}&=& \left|M^{q, {\rm SM}[{\rm NP}]}_{12}\right| e^{2i\bar\beta_q[\theta^{{\rm NP}}_{q}]}\,,\ \   \Gamma^{q, {\rm SM}}_{12}=\left |\Gamma^{q, {\rm SM}[{\rm NP}]}_{12}\right| e^{i\gamma^{{\rm SM}[{\rm NP}]}_q}\,, \non \\
 \Delta^M_q  &=& \left| 1 + r^M_q e^{2 i (\theta^{{\rm NP}}_q -\bar\beta_q)}\right|\,, \ \  r^M_q = \frac{|M^{q,{\rm NP}}_{12}|}{|M^{q,{\rm SM}}_{12}|}\,, \non \\
\Delta^\Gamma_q  &=& \left| 1 + r^\Gamma_q e^{i(\ga^{{\rm NP}}_q -\ga^{\rm SM}_q)}\right|\,, \ \  r^\Gamma_q = \frac{|\Gamma^{q,{\rm NP}}_{12}|}{|\Gamma^{q,{\rm SM}}_{12}|}\,, \non \\
\tan\phi^\Delta_q &=& \frac{r^M_q \sin2(\theta^{{\rm NP}}_{q} -\bar\beta_q)}{1 + r^M_q \cos2(\theta^{{\rm NP}}_{q} -\bar\beta_q)}\,, \ \
\tan\ga^\Delta_q = \frac{ r^\Gamma_q \sin(\ga^{{\rm NP}}_{q} -\ga^{\rm SM}_{q})}{1 - r^\Gamma_q \cos(\ga^{{\rm NP}}_{q} -\ga^{\rm SM}_{q})}\,.
  \ed
Here, the SM contribution is
  \be
  M^{q,SM}_{12}&=&\frac{G^{2}_{F} m^2_{W}}{12 \pi^2}
\eta_{B} m_{B_q}f^{2}_{B_q}\hat{B}_q ( V^*_{tq}V_{tb})^2 S_{0}(x_t) ~,
  \ed
with $S_{0}(x_t)=0.784 x_t^{0.76}$, $x_{t}=(m_t/m_W)^2$ and $\eta_{B}\approx 0.55$ being the QCD correction to $S_0(x_t)$.  The phases appearing in Eq.~(\ref{eq:MG}) are CP-violating phases.  Note that $\bar\beta_d=\beta_d$ and $\bar\beta_s=-\beta_s$.  Using $\phi_q =
\mbox{arg}(-M^q_{12}/\Gamma^q_{12})$, the wrong-sign CA in Eq.~(\ref{eq:aqsl}) with new physics effects on $\Gamma^q_{12}$ and $M^q_{12}$ can be derived as
 \be
 a^q_{s\ell} &=&\frac{\Delta^\Gamma_q}{\Delta^M_q} \frac{\sin\phi_q}{\sin\phi^{\rm SM}_{q}}a^{q}_{s\ell}({\rm SM}) \label{eq:aqsl_new}
 \ed
with $\phi^{\rm SM}_q=2\bar\beta_q -\ga^{\rm SM}_{q}$ and $\phi_q=\phi^{\rm SM}_{q} + \phi^{\Delta}_{q}-\ga^{\Delta}_{q}$.
Furthermore, the mass and rate differences between the heavy and light $B$ mesons are given by
 \be
 \Delta m_{B_q}&=& 2|M^q_{12}|\,,\non\\
 \Delta\Gamma^q &=&\Gamma_L-\Gamma_H= 2 |\Gamma^q_{12}| \cos\phi_q\,. \label{eq:delga}
 \ed

As a comparison, we consider the new physics effect on $K-\bar K$ mixing due to the box diagram with the top quark and the charged Higgs boson in the loop.  This is seen to be the major contribution as other diagrams involving lighter quarks are GIM suppressed in the massless limit or are smaller even when mass effects are taken into account.  The result of $M_{12}$ for the diagram with both the intermediate bosons being the charged Higgs boson is
 \be%
M^{K, HH}_{12}&\approx & \frac{m_K f^2_K (V^*_{td} V_{ts})^2}{12(4\pi)^2 m^2_H} \left\{ \left(|\eta^D|^4 + |\eta^U|^4 -\frac{5}{2} |\eta^D|^2 |\eta^U|^2\right) I_3(y_t) \right. \non \\
&+& 4 y_t J_3(y_t) \left[ \frac{8}{5} \left( \frac{m_K}{m_s+m_d}\right)^2 Re(\eta^{D^*} \eta^U)^2\right. \non \\
&-& \left. \left.  \left( \frac{1}{8}+ \frac{3}{4}\left(\frac{m_K}{m_s +m_d}\right)^2  \right)|\eta^D|^2 |\eta^U|^2 \right]\right\} ~. \label{eq:dk2}
 \ed
The contribution from the diagram with one $W$ boson and one charged Higgs boson in the loop is 
  \be
M^{K,WH}_{12}&\approx & 
\frac{G_F f^2_K m_K}{12\sqrt{2}\pi^2} \left( V^*_{td} V_{ts}\right)^2\non \\
&\times & \left[ - \left( \frac{1}{4} + \frac{3}{2} \left(\frac{m_K}{m_s +m_d}\right)^2\right) \left|\eta^D \right|^2 K_1 (x_t, x_H)+ \frac{m^2_t}{m^2_W}  \left| \eta^U\right|^2 K_2(x_t, x_H) \right] \non ~,
\label{eq:dk2WH}  
  \ed
where $x_t=m^2_t/m^2_W$, $x_H=m^2_{H^\pm}/m^2_W$ and
 \be
 K_n(a,b)= \int^{1}_{0} dx_1 \int^{x_1}_{0} dx_2 \frac{x_2}{(1+(b-1)x_1+ (a-b)x_2)^n}~.
 \ed
  Hadronic effects have been included in Eqs.~(\ref{eq:dk2},\ref{eq:dk2WH}) already.

\end{appendix}

\end{document}